\documentclass[noshowpacs, amsmath, amssymb, amsthm, aps, jmp, thmsb, preprint,11pt]{revtex4}
\usepackage{color,amsbsy,amssymb,latexsym,amsfonts, amsmath,cancel}
\usepackage{mathrsfs}
\usepackage{graphicx}
\usepackage{subfigure}

\begin{document}

\title{Weak Value, Quasiprobability and Bohmian Mechanics}

\author{Kazuki Fukuda$^{1, 2}$}
\email[Email:]{fukudak@post.kek.jp}
\author{Jaeha Lee$^{1, 3}$}
\email[Email:]{jlee@post.kek.jp}
\author{Izumi Tsutsui$^{1, 2, 3}$}
\email[Email:]{izumi.tsutsui@kek.jp}

\affiliation{%
$^1$Theory Center, Institute of Particle and Nuclear Studies, High Energy Accelerator Research Organization (KEK), Tsukuba 305-0801, Japan\\
$^2$School of High Energy Accelerator Science, The Graduate University for Advanced Studies (SOKENDAI), Tsukuba 305-0801, Japan\\
$^3$Department of Physics, University of Tokyo, Tokyo 113-0033, Japan
}%

\date{\today}

\begin{abstract}
We clarify the significance of quasiprobability (QP) in quantum mechanics that is relevant in describing physical quantities associated with a transition process.  
Our basic quantity is Aharonov's weak value, from which the QP can be defined up to a certain ambiguity parameterized by a complex number.  
Unlike the conventional probability, the QP allows us to treat two noncommuting observables consistently, and this is utilized to embed
the QP in Bohmian mechanics such that its equivalence to quantum mechanics becomes more transparent.   We also show that,
with the help of the QP, Bohmian mechanics can be recognized as an ontological
model with a certain type of contextuality.
\end{abstract}

\maketitle

\section{introduction}

Probability is a cornerstone of quantum theory, making its radical departure from classical theory 
which is deterministic in principle.  The probabilistic aspect of quantum theory arises in the form of the Born rule, 
which tells us how probable a particular outcome of measurement is, given a state of the system and an 
observable to be measured.  In the conventional interpretation, 
this probability is directly linked to the relative frequency
of the particular outcome among all possible outcomes in the measurement and, as such, it lies within the range
of zero to one.  This range remains unchanged even if one adopts the Bayesian interpretation of probability advocated recently in quantum mechanics~\cite{Mermin, Fuchs}.

This may no longer be the case when one looks into a process of transition from one (initial) state to another (final) state.
In fact, the notion of transition process is arguably the most foundational element of quantum theory in which a probability amplitude assigned to each process
forms the basic building block for determining the rate of transition, whether it is described in the form of wave function or path integration.  
During a given process, one may
consider the value of a physical observable as a function of the two states which specify the process, and this leads to the so-called weak value
advocated by Aharonov {\it et al.}~\cite{Aharonov1}.  
Despite being generically complex, the weak value has been measured in a number of
systems via the weak measurement, which has now been used for applications including 
precision measurement and direct measurement of quantum states (for a recent review, see, {\it e.g.},\cite{Dressel}).  

The characteristics of the weak
value become particularly acute when the observable is a projection, in which case the conventional expectation value takes a value
in the usual range between zero and one allowing for the probabilistic interpretation.  In contrast, the weak value of the projection
may take a complex value, as has recently been vindicated experimentally \cite{Hasegawa}.   The question posed for us is then
whether the weak value of the projection 
can really be interpreted as something analogous to probability and, if so, how.

The present paper is an attempt to give an answer to this question both from the structural and the conceptual point of view.
Specifically, we clarify how the complex analogue of probability -- occasionally called \lq quasiprobability\rq\ (or \lq pseudo-probability\rq) -- appears naturally in the context
of the weak value when one considers the outcome of measurement.  We shall find that the resultant quasiprobability (QP) 
possesses an intrinsic ambiguity expressed by a complex number $\alpha \in \mathbb{C}$.  Starting from the conditional QP
assigned in the given process, we first furnish both the joint QP and marginal QP according to the standard
procedure of probability theory in a consistent manner.  The weak value arises as an average value with respect to the conditional QP, 
much like the expectation value arising as an average value with respect to the conditional probability. 

This result is then brought to Bohmian mechanics (or de Broglie-Bohm theory), which is presumably the most familiar model of hidden variable theories that
can fully reproduce the outcomes of quantum mechanics, 
where one sees that the physical quantity assigned to an observable takes the form of the weak value.  Interestingly, in terms of the
QP, one can reformulate Bohmian mechanics without changing its content such that the Born rule can be derived
directly, thereby rendering the equivalence to quantum mechanics in treasting measurement outcomes more transparent.

The conceptual aspect of QP is then studied further in a more general setting of ontological models, where we
show that Bohmian mechanics \cite{Bohm1,Bohm2} can be regarded as an ontological model with a certain type of contextuality.  
This provides an arena where one can examine the validity of QP and at the same time see how the ambiguity represented
by the parameter $\alpha$ disappears in physical quantities obtained experimentally.

This paper is organized as follows.  After the Introduction, 
in section 2 we recall how the weak value calls for the notion of QP in quantum mechanics in a given transition process.  From the 
conditional QP defined there follows the 
joint QP as well as the marginal one.    
In section 3, by reconsidering Bohmian mechanics based on 
the QP, we exhibit the significance of the weak value and also present
the Born rule as a direct consequence of the premises we revise.  
The fact that Bohmian mechanics falls into the category of ontological models is discussed in section 4, where the nonlocality of Bohmian mechanics is addressed with respect to the conventional formalism of hidden variable theories.
Finally, section 5 is devoted to our conclusion and discussion.

\section{quasiprobability underlying the Weak Value}

We begin our discussion by observing how QP emerges in quantum mechanics when a transition process is considered.
Specifically, the QP we shall find is conditioned by a pair of states which specifies the transition, and from this conditional QP both the joint QP and the marginal QP are derived according to the standard procedure of probability theory.  It will be seen that our conditional QP fulfills (an extended version of) the Kolmogorov axioms and shares some, though not all, common features with the conventional probability.

\subsection{Conditional QP Distribution}

We first recall that, given two states $|\psi\rangle$ and  $|\phi\rangle$ of a Hilbert space $\mathcal{H}$ with $\left\langle \phi|\psi\right\rangle \neq0$, 
the weak value of an observable $A$ is defined by
\begin{equation}
A_{\mathrm{w}} =\frac{\left\langle \phi\right|A\left|\psi\right\rangle }{\left\langle \phi|\psi\right\rangle }.
\label{eq:weak value}
\end{equation}
Despite its complex-valuedness, the weak value can actually be measured
by the process called \lq weak measurement\rq\ \cite{Aharonov1}.   It has thus become important for us to investigate the conceptual 
significance of the weak value regarding its physical reality, and for this we explore below the implication of the weak value in the 
extension of probability in quantum mechanics.

To this end, let us first consider the spectrum decomposition of the observable
$A=\int a\, E^{A}\left(a\right)\mathrm{d}a$, where $E^{A}\left(a\right)=\left|a\right\rangle \left\langle a\right|$ is the projector
associated with the eigenstate 
$A \left|a\right\rangle = a\left|a\right\rangle$ of $A$.   
With this decomposition, 
the weak value (\ref{eq:weak value}) 
may be written as
\begin{equation}
\label{eq:wvexp}
A_{\mathrm{w}}
=\int a\frac{\left\langle \phi\right|E^{A}\left(a\right)\left|\psi\right\rangle }{\left\langle \phi|\psi\right\rangle }\mathrm{d}a
= \int a \,p\left(a\left|\psi,\phi\right.\right)\mathrm{d}a,
\end{equation}
in terms of the weak value of the projection operator,
\begin{equation}
p\left(a\left|\psi,\phi\right.\right) = \frac{\left\langle \phi\right|E^{A}\left(a\right)\left|\psi\right\rangle }{\left\langle \phi|\psi\right\rangle }.
\label{eq:qpdef}
\end{equation}
The expression (\ref{eq:wvexp}) suggests that the function $a\mapsto p\left(a\left|\psi,\phi\right.\right)$ may be interpreted as an analogue of
probability in the sense that its average yields the weak value, despite that the value $p\left(a\left|\psi,\phi\right.\right)$ may go beyond the
standard range $[0, 1]$ or even becomes complex.  
Such an extended 
probability provides an example of the QP, but before we mention its general properties, we 
briefly sketch how it appears  
in quantum mechanics when a process of transition is considered.

Let $\mathcal{P}$ be the set of all projection operators
on a (finite dimensional) Hilbert space $\mathcal{H}$.  We denote by 
$P_{|\chi\rangle} = |\chi\rangle \langle \chi| \in \mathcal{P}$ 
the projection operator associated with a state $|\chi\rangle \in \mathcal{H}$, and assume hereafter that all states are normalized.
Now, let $f_{\psi,\phi}:\mathcal{P}\to\mathbb{C}$ be a
map from $\mathcal{P}$ to the complex plane $\mathbb{C}$ for a given pair of states, $|\psi\rangle, |\phi\rangle \in\mathcal{H}$ for which $\left\langle \phi|\psi\right\rangle \neq0$.  
We demand that 
the map $f_{\psi,\phi}$ fulfill the following conditions:
\begin{enumerate}
\item[C1)] 
For $P_{i}, P_{j} \in \mathcal{P}$ which are mutually
orthogonal $P_{i}P_{j}=0$, $i\neq j$, we have
\begin{equation}
f_{\psi,\phi}\left(\sum_{i}P_{i}\right)=\sum_{i}f_{\psi,\phi}\left(P_{i}\right).
\end{equation}
\item[C2)] 
Let $\left|\chi^{\perp}\right\rangle$ be a state orthogonal to $|\chi\rangle$, that is, $\left\langle \chi^{\perp}|\chi\right\rangle =0$.  
Then we have
\begin{equation}
f_{\psi,\phi}\left(P_{|\psi\rangle}\right)=1, \quad
f_{\psi,\phi}\left(P_{|\psi^{\perp}\rangle}\right)=0, \quad
f_{\psi,\phi}\left(P_{|\phi\rangle}\right)=1, \quad
f_{\psi,\phi}\left(P_{|\phi^{\perp}\rangle}\right)=0.
\end{equation}
\end{enumerate}

It has been shown by Morita {\it et al.}~\cite{Morita} that, for $\mathrm{dim}\, \mathcal{H}\geq3$, any map $f_{\psi,\phi}$ that obeys conditions C1 and C2
must be of the form,
\begin{equation}
f_{\psi,\phi}\left(P_i\right)=\alpha\frac{\left\langle \phi\right|P_i\left|\psi\right\rangle }{\left\langle \phi|\psi\right\rangle }+\left(1-\alpha\right)\frac{\left\langle \psi\right|P_i\left|\phi\right\rangle }{\left\langle \psi|\phi\right\rangle }
\label{eq:morita-1}
\end{equation}
for some $\alpha\in\mathbb{C}$.  We mention that condition C1 is tantamount to one of the conditions used in Gleason's
theorem \cite{Gleason} to derive the Born rule and condition C2 is also analogous to the other one used there, or more specifically, C2 imposes consistency for two states while the counterpart in \cite{Gleason} does it for one state.   Indeed, as we shall see shortly, when the two states are identical $|\psi\rangle = |\phi\rangle$, the form (\ref{eq:morita-1}) reduces to the
expectation value $\left\langle \psi\right|P_i\left|\psi\right\rangle$ leading to the Born rule.

In view of the generality admitted to the map (\ref{eq:morita-1}), let us extend the QP in (\ref{eq:qpdef}) with the parameter $\alpha$ and consider
\begin{equation}
p^{\alpha}\left(a\left|\psi,\phi\right.\right):=\alpha\frac{\left\langle \phi\right|E^{A}\left(a\right)\left|\psi\right\rangle }{\left\langle \phi|\psi\right\rangle }+\left(1-\alpha\right)\frac{\left\langle \psi\right|E^{A}\left(a\right)\left|\phi\right\rangle }{\left\langle \psi|\phi\right\rangle }.
\label{eq:eq_of_def_cqp_al}
\end{equation}
For the reason that will soon become apparent, we call $p^{\alpha}\left(a\left|\psi,\phi\right.\right)$ the
{\it $\alpha$-parameterized conditional QP distribution}
of an observable $A$. 

One of the advantages of this $\alpha$-extension is that we can then tune the mixture of the processes associated with the QP
in (\ref{eq:eq_of_def_cqp_al}) freely by choosing the value of $\alpha$ properly.
For example, by choosing $\alpha=1$, we have 
\begin{equation}
p^{\alpha=1}\left(a\left|\psi,\phi\right.\right)=\frac{\left\langle \phi\right|E^{A}\left(a\right)\left|\psi\right\rangle }{\left\langle \phi|\psi\right\rangle },
\end{equation}
which is our original one (\ref{eq:qpdef}) defined for the process $|\psi\rangle \to |\phi\rangle$,
while by choosing $\alpha=0$, we find 
\begin{equation}
p^{\alpha=0}\left(a\left|\psi,\phi\right.\right)=\frac{\left\langle \psi\right|E^{A}\left(a\right)\left|\phi\right\rangle }{\left\langle \psi|\phi\right\rangle }=p^{\alpha=1}\left(a\left|\psi,\phi\right.\right)^{*}
\end{equation}
which is the one for the reverse process $|\phi\rangle \to |\psi\rangle$. 
The intermediate choice $\alpha=1/2$ then gives the equal mixture of the two, which amounts to taking the real part of the original one,
\begin{eqnarray}
p^{\alpha=1/2}\left(a\left|\psi,\phi\right.\right) 
& = & \frac{1}{2}\left\{ \frac{\left\langle \phi\right|E^{A}\left(a\right)\left|\psi\right\rangle }{\left\langle \phi|\psi\right\rangle }+\frac{\left\langle \psi\right|E^{A}\left(a\right)\left|\phi\right\rangle }{\left\langle \psi|\phi\right\rangle }\right\} \nonumber \\
 & = & \mathrm{Re}\frac{\left\langle \phi\right|E^{A}\left(a\right)\left|\psi\right\rangle }{\left\langle \phi|\psi\right\rangle }.
\end{eqnarray}

In particular, for a closed process $|\psi\rangle \to |\psi\rangle$, one observes that the $\alpha$-dependence disappears to yield
\begin{equation}
p^{\alpha}\left(a\left|\psi,\psi\right.\right) 
=\left\langle \psi\right|E^{A}\left(a\right)\left|\psi\right\rangle =\left|\left\langle a|\psi\right\rangle \right|^{2}.
\label{eq:psi=00003Dphi conditional qp}
\end{equation}
We thus notice that, if one is allowed to interpret that a closed process is equivalent to adopting no further condition other than the initial state $|\psi\rangle$
allowing for the identification  $p^{\alpha}\left(a\left|\psi,\psi\right.\right)  = p\left(a\left|\psi\right.\right)$ where $p\left(a\left|\psi\right.\right)$ represents 
the probability of finding the value $a$ when the state is $|\psi\rangle$, one obtains
$p\left(a\left|\psi\right.\right) = \left|\left\langle a|\psi\right\rangle \right|^{2}$, namely, the Born rule.  Although we do not actually adopt this interpretation in this paper, the connection with the Born rule suggests that our QP is somehow consistent with the standard notion of probability in quantum mechanics.
 
\subsection{Joint QP Distribution}

Next, we proceed to define the joint QP and the marginal QP from the conditional QP introduced above.
Let $B$ be an observable which we use as a \lq reference\rq\ in providing the joint QP, 
and let $\{ b\}$ be the set of eigenvalues of $B$.  
The observable $B$ is chosen independently from $A$ and hence it 
does not commute with $A$ in general.  
As before we denote by $p\left(b\left|\psi\right.\right)$
the probability 
that the observable $B$ takes the value $b$ when the state is $|\psi\rangle$.
Then, following the standard procedure of probability theory, we may define the \emph{$\alpha$-parameterized}
\emph{joint QP distribution} as
\begin{equation}
p^{\alpha}\left(b,a\left|\psi\right.\right):=p^{\alpha}\left(a\left|\psi,b\right.\right)p\left(b\left|\psi\right.\right),
\label{eq:def_of_jointQuasipro}
\end{equation}
from the conditional QP in (\ref{eq:eq_of_def_cqp_al}) adopted in the present situation, that is,
\begin{equation}
p^{\alpha}\left(a\left|\psi,b\right.\right) = 
\alpha\frac{\left\langle  b\, \right|E^{A}\left(a\right)\left|\psi\right\rangle }{\left\langle b\, |\psi\right\rangle }+\left(1-\alpha\right)\frac{\left\langle \psi\right|E^{A}\left(a\right)\left| b \right\rangle }{\left\langle \psi| b \right\rangle }.
\label{eq:def_conprob}
\end{equation}
Here we make an assumption that for the reference observable $B$ we have the probability formula,
\begin{equation}
\label{eq:born}
p\left(b\left|\psi\right.\right) = \left|\left\langle b|\psi\right\rangle \right|^{2}.
\end{equation}
This allows us to obtain
\begin{eqnarray}
p^{\alpha}\left(b,a\left|\psi\right.\right) 
& = & \alpha\frac{\left\langle b \right|E^{A}\left(a\right)\left|\psi\right\rangle }{\left\langle b|\psi\right\rangle }\left|\left\langle b|\psi\right\rangle \right|^{2} +\left(1-\alpha\right)\frac{\left\langle \psi\right|E^{A}\left(a\right)\left|b\right\rangle }{\left\langle \psi|b\right\rangle }\left|\left\langle b|\psi\right\rangle \right|^{2}\nonumber \\
& = & \alpha\left\langle \psi\right|E^{B}\left(b\right)E^{A}\left(a\right)\left|\psi\right\rangle +\left(1-\alpha\right)\left\langle \psi\right|E^{A}\left(a\right)E^{B}\left(b\right)\left|\psi\right\rangle,
\label{eq:joint quasi-pro.-1}
\end{eqnarray}
where $E^{B}\left(b\right)=\left|b\right\rangle \left\langle b \right|$ is the projection onto the eigenspace of the eigenvalue $b$.
This suggests that 
the $\alpha$-parameterized joint QP distribution may
be regarded as the generalization of the Kirkwood-Dirac function \cite{Kirkwood, Dirac-Quasi}
or the weak joint QP argued by Ozawa \cite{Ozawa_QuasiProb}.
It should be noted that the joint QP distribution $p^{\alpha}\left(b,a\left|\psi\right.\right)$
is not invariant $p^{\alpha}\left(b,a\left|\psi\right.\right)\neq p^{\alpha}\left(a,b\left|\psi\right.\right)$ under the interchange of $E^{A}\left(a\right)$ and
$E^{B}\left(b\right)$, and that our use of the probability formula (\ref{eq:born}) to derive 
the joint QP in (\ref{eq:joint quasi-pro.-1}) is restricted to the particular reference observable $B$ we have chosen.

In order to investigate the properties of the $\alpha$-parameterized
joint QP,
it is convenient for us to introduce the $\circ_{\alpha}$-product for two operators $X$ and $Y$
on $\mathcal{H}$ as
\begin{equation}
X\circ_{\alpha}Y:=\alpha XY+\left(1-\alpha\right)YX,
\label{eq:alphaproduct}
\end{equation}
with which the joint QP becomes 
\begin{equation}
p^{\alpha}\left(b,a\left|\psi\right.\right)=\left\langle \psi\right|E^{B}\left(b\right)\circ_{\alpha}E^{A}\left(a\right)\left|\psi\right\rangle.
\label{eq:jqp}
\end{equation}
One then confirms at once that the $\alpha$-parameterized joint QP
distribution with $\alpha=1/2$ is real-valued and invariant
under the interchange of $E^{A}\left(a\right)$ and $E^{B}\left(b\right)$,
\begin{eqnarray}
p^{\alpha=1/2}\left(a,b\left|\psi\right.\right) & = & \mathrm{Re}\left\langle \psi\right|E^{B}\left(b\right)E^{A}\left(a\right)\left|\psi\right\rangle \nonumber \\
 & = & \mathrm{Re}\left\langle \psi\right|E^{A}\left(a\right)E^{B}\left(b\right)\left|\psi\right\rangle =p^{\alpha=1/2}\left(b,a\left|\psi\right.\right).
\end{eqnarray}

Moreover, one can observe (see (\ref{eq:commutatorrrrr}) in the Appendix) that
\begin{equation}
p^{\alpha}\left(b,a\left|\psi\right.\right)-p^{\alpha}\left(a,b\left|\psi\right.\right)=\left(2\alpha-1\right)\left\langle \psi\right|\left[E^{B}\left(b\right),E^{A}\left(a\right)\right]\left|\psi\right\rangle,
\end{equation}
which shows that, if $A$ and $B$ commute, then the joint QP $p^{\alpha}\left(b,a\left|\psi\right.\right)=p^{\alpha}\left(a,b\left|\psi\right.\right)$ becomes 
$\alpha$-independent and reads
\begin{equation}
p^{\alpha}\left(b,a\left|\psi\right.\right)=\left\langle \psi\right|E^{B}\left(b\right)E^{A}\left(a\right)\left|\psi\right\rangle 
=\left\langle \psi\right|E^{A}\left(a\right)E^{B}\left(b\right)\left|\psi\right\rangle 
\label{eq:compatible joint quasi-prob}
\end{equation}
for any $\alpha\in\mathbb{C}$. 
In fact, when the two observables commute, the conditional QP in (\ref{eq:def_conprob}) is already  $\alpha$-independent and simplifies as
\begin{equation}
p^{\alpha}\left(a\left|\psi, b\right.\right) = 
\frac{\left\langle  b\, \right|E^{A}\left(a\right)\left|\psi\right\rangle }{\left\langle b\, |\psi\right\rangle }
= \frac{\left\langle \psi\right|E^{A}\left(a\right)\left| b \right\rangle }{\left\langle \psi| b \right\rangle }.
\end{equation}

\subsection{Marginal QP Distribution}

Finally, we shall define the marginal QP distribution of
$A$ on $\psi$ by analogy with the usual joint probability,
\begin{equation}
p^{\alpha}\left(a\left|\psi\right.\right):=\int p^{\alpha}\left(b,a\left|\psi\right.\right)\mathrm{d}b
=\int p^{\alpha}\left(a\left|\psi,b\right.\right)p\left(b\left|\psi\right.\right)\mathrm{d}b.
\label{eq:def_of_marginal_alpha}
\end{equation}
We then find from (\ref{eq:jqp}) and (\ref{eq:xident}) that this marginal QP distribution
is independent of the parameter $\alpha$ and reduces to the conventional probability of obtaining a value $a$ of $A$,
\begin{eqnarray}
p^{\alpha}\left(a\left|\psi\right.\right)
 & = & \int\left\langle \psi\right|E^{B}\left(b\right)\circ_{\alpha}E^{A}\left(a\right)\left|\psi\right\rangle \mathrm{d}b\nonumber \\
 & = & \left\langle \psi\right|E^{A}\left(a\right)\left|\psi\right\rangle \nonumber \\
 & = & \left|\left\langle a|\psi\right\rangle \right|^{2}.
\label{eq:reproduction of Born rule}
\end{eqnarray}
Since the choice of $A$ is arbitrary, 
the result (\ref{eq:reproduction of Born rule}) shows that, whatever the interpretation one attaches to the 
$\alpha$-parameterized conditional QP distribution (\ref{eq:eq_of_def_cqp_al}), one ends up with
the conventional Born rule at the level of the marginal distribution that can be directly tested by measurement.

At this point, we remark that (\ref{eq:reproduction of Born rule}) together with (\ref{eq:def_of_marginal_alpha}) resembles the reproduction
condition in the ontological model considered in \cite{ontological_model_original}, if the eigenstate $\left|b\right\rangle $
of the observable $B$ is regarded as an \emph{ontic state}.  However, this is not quite the case since our QP is $\psi$-dependent in general while it is not in the ontological model.   More on this will come later when we discuss the relevance of QP in the ontological model.

The foregoing discussions indicate that our ($\alpha$-parameterized) conditional QP in
(\ref{eq:eq_of_def_cqp_al}) provides a key ingredient of quantum theory in that it yields the Born rule as marginal distribution.  
The assumption we adopted to achieve this is the probability formula (\ref{eq:born}) for the reference observable $B$, and 
at this point, we mention that, because of this special status of input, the reference observable $B$ acquires a deterministic
property in the sense that
\begin{equation}
\label{eq:deterministic}
p^{\alpha}\left(b'\left|\psi,b\right.\right)
=\begin{cases}
1, & \hbox{for} \quad b=b',\\
0, & \hbox{for} \quad b\neq b.
\end{cases}
\end{equation}
When the reference observable $B$ is chosen to be the position observable $X$, 
our observation implies that the position becomes a deterministic variable, as will be seen in Bohmian mechanics in the next section.

\subsection{General Description of QP}

Before closing this section, we give a brief account of the mathematical description of QP in a form slightly more general than (\ref{eq:eq_of_def_cqp_al}).
Let us denote by $\Delta \subset \mathbb{R}$ a collection of outcomes for the measurement of an observable $A$.
The $\alpha$-parameterized conditional QP associated with $\Delta$ is then written as
\begin{equation}
p^{\alpha}\left(a\in\Delta\left|\psi,\phi\right.\right)=\int_{\Delta}p^{\alpha}\left(a\left|\psi,\phi\right.\right)\mathrm{d}a.
\end{equation}
From (\ref{eq:eq_of_def_cqp_al}) we learn that it satisfies 
\begin{equation}
p^{\alpha}\left(a\in\Delta\left|\psi,\phi\right.\right)=\alpha\frac{\left\langle \phi\right|E^{A}\left(\Delta\right)\left|\psi\right\rangle }{\left\langle \phi|\psi\right\rangle }+\left(1-\alpha\right)\frac{\left\langle \psi\right|E^{A}\left(\Delta\right)\left|\phi\right\rangle }{\left\langle \psi|\phi\right\rangle },
\end{equation}
where $E^{A}\left(\Delta\right)$ is the projection onto the subspace spanned by the eigenstates associated with $\Delta$.
The QP distribution defined above qualifies as a complex-valued version of the Kolmogorov axioms of the probability
measure, namely,
\begin{enumerate}
\item[K1)] 
Countable additivity:
\begin{equation}
p^{\alpha}\left(a\in\Delta\left|\psi,\phi\right.\right)=\sum_{i}p^{\alpha}\left(a\in\Delta_{i}\left|\psi,\phi\right.\right)
\end{equation}
for any mutually disjoint sequence of intervals $\Delta_{1},\Delta_{2},\ldots$
with $\Delta=\cup_{i}\Delta_{i}$, and
\item[K2)] 
Normalization condition:
\begin{equation}
p^{\alpha}\left(a\in\mathbb{R}\left|\psi,\phi\right.\right)=1.
\end{equation}
\end{enumerate}
Indeed, for K1 we observe
\begin{eqnarray}
p^{\alpha}\left(a\in\Delta\left|\psi,\phi\right.\right) & = & \alpha\frac{\left\langle \phi\right|\sum_{i}E^{A}\left(\Delta_{i}\right)\left|\psi\right\rangle }{\left\langle \phi|\psi\right\rangle }+\left(1-\alpha\right)\frac{\left\langle \psi\right|\sum_{i}E^{A}\left(\Delta_{i}\right)\left|\phi\right\rangle }{\left\langle \psi|\phi\right\rangle }
\nonumber \\
 & = & \sum_{i}\left\{ \alpha\frac{\left\langle \phi\right|E^{A}\left(\Delta_{i}\right)\left|\psi\right\rangle }{\left\langle \phi|\psi\right\rangle }+\left(1-\alpha\right)\frac{\left\langle \psi\right|E^{A}\left(\Delta_{i}\right)\left|\phi\right\rangle }{\left\langle \psi|\phi\right\rangle }\right\} 
\nonumber \\
 & = & \sum_{i}p^{\alpha}\left(a\in\Delta_{i}\left|\psi,\phi\right.\right),
\end{eqnarray}
and for K2 we have
\begin{eqnarray}
p^{\alpha}\left(a\in\mathbb{R}\left|\psi,\phi\right.\right) = \alpha\frac{\left\langle \phi\right|E^{A}\left(\mathbb{R}\right)\left|\psi\right\rangle }{\left\langle \phi|\psi\right\rangle }+\left(1-\alpha\right)\frac{\left\langle \psi\right|E^{A}\left(\mathbb{R}\right)\left|\phi\right\rangle }{\left\langle \psi|\phi\right\rangle }
= 1,
\end{eqnarray}
since $E^{A}\left(\mathbb{R}\right) = I$.

Those maps satisfying K1 and K2 are termed \emph{QP measure} or \emph{complex
probability measure}.  The difference between the
QP and the conventional probability is simply that the value of the latter is restricted to the real range $[0,1]$ whereas the former is allowed to take any complex numbers. 

\section{quasiprobability and Bohmian mechanics}

We have so far argued that the QP may be regarded as a fundamental ingredient of quantum mechanics leading to the Born rule.
We now show that the QP can also be found in the most familiar type of realistic interpretations of quantum mechanics, namely, 
Bohmian mechanics.  
In particular, with the QP we can safely say that 
Bohmian mechanics is a contextual version of the ontological model proposed by Spekkens \cite{Spekkens_Contextuality, ontological_model_original}.

\subsection{Bohmian Mechanics and the Local Expectation Value}

To confirm the above statements, we first recall the framework of Bohmian mechanics \cite{Bohm1, Bohm2}. 
The basic postulates of Bohmian mechanics, considered here for a nonrelativistic scalar particle with mass $m$ for simplicity (the extension to the multi-particle or non-scalar case can be made straightforwardly), may be stated as follows:
\begin{enumerate}
\item[B1)] 
The state of the particle is described partly by 
a vector $|\psi\rangle$ in a Hilbert space $\mathcal{H}$ which obeys the Schr{\" o}dinger equation,
\begin{equation}
i\hbar\frac{\partial}{\partial t}|\psi\rangle = H |\psi\rangle,
\label{eq:scheq}
\end{equation}
with $H$ being a self-adjoint operator called \lq Hamiltonian\rq. 
\item [B2)]  
The position $\mathbf{x} \in \mathbb{R}^{3}$ of the particle in the state $|\psi\rangle$ is distributed randomly according to the 
probability,
\begin{equation}
p(\mathbf{x}\left|\psi\right.)=\left|\psi(\mathbf{x})\right|^{2},
\label{eq:ignorance}
\end{equation}
where $\psi\left(\mathbf{x}\right)=\langle \mathbf{x}|\psi\rangle$ is the position representation of the state called \lq wave function\rq.
\item [B3)] 
Given a position $\mathbf{x}$ of the particle in the state $|\psi\rangle$, 
the momentum $\mathbf{p}= m (d\mathbf{x}/dt)$ of the particle in the state $|\psi\rangle$ is determined by
\begin{equation}
\mathbf{p} =  \hbar\, \mathrm{Im}\frac{\nabla\psi\left(\mathbf{x}\right)}{\psi\left(\mathbf{x}\right)}.
\label{eq:def.of.momentum}
\end{equation}
\end{enumerate}

Note that, like in classical mechanics, in Bohmian mechanics the dynamical evolution of the particle $\mathbf{x}(t)$
is completely determined from (\ref{eq:scheq}) and (\ref{eq:def.of.momentum}), 
once the initial data, $\mathbf{x}(0)$ and $\mathbf{p}(0)$, are provided.
In this sense, Bohmian mechanics is
deterministic, even though it is statistical on account of the random distribution of the position (\ref{eq:ignorance}).
However, the price Bohmian mechanics pays for it is that, 
compared to the standard quantum mechanics, it requires to use 
the position $\mathbf{x}$ of the particle as an additional variable, which plays the role of the so-called \lq hidden variable\rq\
even though it can be measured (what is hidden instead is the state $|\psi\rangle$ which is not subject to direct measurement).

Note also that the wave
function $\psi(\mathbf{x})$ plays a double role, {\it i.e.}, it gives the probability distribution (\ref{eq:ignorance}) and, at the same time, 
determines the dynamical development of the position by (\ref{eq:def.of.momentum}).   Because of the latter role, the wave function is
sometimes called the {\it guiding wave}.
It should be mentioned that Bohmian mechanics
does not contain the Born rule in the premises, and hence one has to argue separately how it can be gained in order to 
reproduce the predictions of quantum theory completely.  This has been done in \cite{Bohm1,Bohm2} by
taking account of the measurement procedure. 

Concerning the representation of an observable $A$ in Bohmian mechanics, Holland \cite{HollandBook} has introduced 
the {\it local expectation value}, 
\begin{equation}
\left\langle A\right\rangle _{\psi}\left(\mathbf{x}\right)
:=\mathrm{Re}\frac{\psi^{*}\left(\mathbf{x}\right)\left(A\psi\right)\left(\mathbf{x}\right)}{\psi^{*}\left(\mathbf{x}\right)\psi\left(\mathbf{x}\right)}
=\mathrm{Re}\frac{\left\langle \mathbf{x}\right|A\left|\psi\right\rangle }{\left\langle \mathbf{x}|\psi\right.\rangle }.
\label{eq:LEV}
\end{equation}
One confirms that the average $\left\langle A\right\rangle _{\psi}$ of the local expectation value
over the probability $p(\mathbf{x}|\psi)$ is equal to the expectation value of quantum mechanics,
\begin{eqnarray}
\left\langle A\right\rangle _{\psi}
&=&  
\int \left\langle A\right\rangle _{\psi}\left(\mathbf{x}\right)p(\mathbf{x}|\psi)\mathrm\, {d}\mathbf{x} \nonumber \\
&=&  \int\mathrm{Re}\frac{\left\langle \mathbf{x}\right|A\left|\psi\right\rangle }{\left\langle \mathbf{x}|\psi\right\rangle }\left|\left\langle \mathbf{x}|\psi\right\rangle \right|^{2}\mathrm{d}\mathbf{x} \nonumber \\
&=&  \left\langle \psi\right|A\left|\psi\right\rangle.
\label{eq:LEVexpec}
\end{eqnarray}

An important point is that the local expectation value (\ref{eq:LEV}) may be regarded as a generalization
of the relation (\ref{eq:def.of.momentum}).  Indeed, recalling that ${-i\hbar\nabla}$ is the representation of the momentum operator $P$ in the position representation, one has
\begin{eqnarray}
\left\langle P\right\rangle _{\psi}\left(\mathbf{x}\right) 
=  \mathrm{Re}\frac{\left\langle \mathbf{x}\right| P \left|\psi\right\rangle }{\left\langle \mathbf{x}|\psi\right\rangle } 
=  \mathrm{Re}\frac{-i\hbar\nabla\psi\left(\mathbf{x}\right)}{\psi\left(\mathbf{x}\right)}.
\label{eq:LEVofMomentum}
\end{eqnarray}
This coincides with (\ref{eq:def.of.momentum}) if one identifies the local expectation value of the momentum observable $P$ with the value of the momentum $\mathbf{p}$ of the particle.
By virtue of (\ref{eq:LEVexpec}) and (\ref{eq:LEVofMomentum}),
the local expectation value of $A$ can be
understood as a physical value associated with $A$, and with this in mind, 
one may replace (\ref{eq:def.of.momentum}) with the more general (\ref{eq:LEV}).  
Obviously, the key element behind this is the fact that the local expectation value is none other than the real part of weak value, 
which has been mentioned earlier in \cite{Wiseman} for the case of momentum (or velocity).

\subsection{Bohmian Mechanics with QP}

We now introduce the $\alpha$-parameterized QP
in Bohmian mechanics, but before doing so, let us consider the position operator
$X$ and the projection operator  $E^{X}\left(\mathbf{x}\right)=\left|\mathbf{x}\right\rangle \left\langle \mathbf{x}\right|$
on the eigenstate $\left|\mathbf{x}\right\rangle $ of $X$. 
Consider then the $\alpha$-parameterized joint QP
distribution (\ref{eq:joint quasi-pro.-1}) in the notation (\ref{eq:jqp}) by choosing the reference observable as $B = X$, that is,
\begin{equation}
p^{\alpha}\left(\mathbf{x},a\left|\psi\right.\right)=\left\langle \psi\right|E^{X}\left(\mathbf{x}\right)\circ_{\alpha}E^{A}\left(a\right)\left|\psi\right\rangle.\label{eq:dBBjointQuasi}
\end{equation}
Recall that the joint QP leads to the marginal QP 
which ensures the probability formula (\ref{eq:reproduction of Born rule}) for any observable $A$, {\it i.e.}, the Born rule.   As mentioned before, this formula is not included in the original postulates B1 - B3 of Bohmian mechanics and hence has to be proven separately.

We also note that, once the joint QP given by (\ref{eq:dBBjointQuasi}) is adopted, the conditional QP distribution
of an observable $A$ is deduced from (\ref{eq:def_of_jointQuasipro}) as
\begin{eqnarray}
p^{\alpha}\left(a\left|\psi,\mathbf{x}\right.\right) & = & \alpha\frac{\left\langle \mathbf{x}\right|E^{A}\left(a\right)\left|\psi\right\rangle }{\left\langle \mathbf{x}|\psi\right\rangle }+\left(1-\alpha\right)\frac{\left\langle \psi\right|E^{A}\left(a\right)\left|\mathbf{x}\right\rangle }{\left\langle \psi|\mathbf{x}\right\rangle }.\label{eq:conditionalquasipro. dBB}
\end{eqnarray}
In particular, the conditional QP distribution (\ref{eq:conditionalquasipro. dBB})
of position $X$ becomes
\begin{eqnarray}
p^{\alpha}\left(\mathbf{x}'\left|\psi,\mathbf{x}\right.\right) 
& = & \alpha\frac{\langle \mathbf{x} |\mathbf{x}'\rangle \langle \mathbf{x}' | \psi\rangle }{\left\langle \mathbf{x}|\psi\right\rangle }
+\left(1-\alpha\right)\frac{\langle \psi |\mathbf{x}'\rangle \langle \mathbf{x}' |\mathbf{x}\rangle }{\left\langle \psi|\mathbf{x}\right\rangle } \nonumber \\
 & = & \left\{ \alpha\frac{\left\langle \mathbf{x}'|\psi\right\rangle }{\left\langle \mathbf{x}|\psi\right\rangle }+\left(1-\alpha\right)\frac{\left\langle \psi|\mathbf{x}'\right\rangle }{\left\langle \psi|\mathbf{x}\right\rangle }\right\} \delta\left(\mathbf{x}-\mathbf{x}'\right).
\end{eqnarray}
This shows that, should the joint QP in (\ref{eq:dBBjointQuasi}) be introduced, the deterministic nature of Bohmian mechanics becomes manifest as mentioned earlier in (\ref{eq:deterministic}) (in the discrete case).

From the $\alpha$-parameterized conditional QP,
the conditional average of $A$ is evaluated as
\begin{eqnarray}
\int a\, p^{\alpha}\left(a\left|\psi, \mathbf{x}\right.\right)\mathrm{d}a
= \alpha\frac{\left\langle \mathbf{x}\right|A\left|\psi\right\rangle }{\left\langle \mathbf{x}|\psi\right\rangle }
 +\left(1-\alpha\right)\frac{\left\langle \psi\right|A\left|\mathbf{x}\right\rangle }{\left\langle \psi|\mathbf{x}\right\rangle }.
\label{eq:extc2}
\end{eqnarray}
We thus observe that Holland's local expectation value $\left\langle A\right\rangle _{\psi}\left(\mathbf{x}\right)$ in (\ref{eq:LEV}) arises as our expectation value at $\alpha=1/2$.
Note that the expectation value (\ref{eq:extc2}) is a linear combination of the conjugate pair of the two weak values (\ref{eq:weak value}) obtained 
by reversing the process of transition.

Now, since (\ref{eq:def.of.momentum}) in postulate B3  can be replaced with (\ref{eq:LEV}), and since (\ref{eq:LEV}) allows for 
the extension (\ref{eq:extc2}), we may just replace B3 with the new postulate:
\begin{description}
\item{B$3^\prime$)}
Given a position $\mathbf{x}$ of the particle in the state $|\psi\rangle$, 
the value of an observable $A$ is given by 
\begin{eqnarray}
\left\langle A\right\rangle_\psi^{\alpha}\left(\mathbf{x}\right) 
:= \alpha\frac{\left\langle \mathbf{x}\right|A\left|\psi\right\rangle }{\left\langle \mathbf{x}|\psi\right\rangle }
 +\left(1-\alpha\right)\frac{\left\langle \psi\right|A\left|\mathbf{x}\right\rangle }{\left\langle \psi|\mathbf{x}\right\rangle },
\label{eq:newvalue}
\end{eqnarray}
with $\alpha \in \mathbb{C}$.
\end{description}

At this point, it is important to recognize that adopting B$3^\prime$ in place of B3 does not modify the content of Bohmian mechanics, since
the $\alpha$-dependence appears only in the association of the value (\ref{eq:newvalue}) but not in the final probability.  
Indeed, as in (\ref{eq:LEVexpec}) the average value has no $\alpha$-dependence and remains the same,
\begin{eqnarray}
\left\langle A\right\rangle^\alpha_{\psi} 
= \int \left\langle A\right\rangle^\alpha_{\psi}
\left(\mathbf{x}\right)p(\mathbf{x}|\psi)
\mathrm{d}\mathbf{x} 
=  \left\langle \psi\right|A\left|\psi\right\rangle,
\label{eq:averagevalue}
\end{eqnarray}
as can be confirmed readily by use of (\ref{eq:ignorance}) in B2.  
Similarly, from (\ref{eq:newvalue}) one can derive (\ref{eq:conditionalquasipro. dBB}) by choosing
$E^A(a)$ for $A$, and again by using  (\ref{eq:ignorance}) one can obtain the joint QP in (\ref{eq:dBBjointQuasi})
from which the Born rule follows directly.  
We thus learn that our QP can be embedded in Bohmian mechanics without altering its physical content.
As shown in (\ref{eq:averagevalue}), the $\alpha$-dependence in (\ref{eq:newvalue}), which exhibits the ambiguity in the value of $A$ at a position $\mathbf{x}$, disappears in the physically measurable quantities after the average over all possible $\mathbf{x}$ is performed.

\section{Bohmian Mechanics and the Ontological Model}

Finally, in order to see a deeper role of QP in Bohmian mechanics, we examine the connection between Bohmian mechanics
and the ontological model of quantum theory introduced by Harrigan
and Spekkens \cite{ontological_model_original}. 
It has been pointed out in \cite{Feintzeig}
that the conventional framework of the ontological model cannot accommodate Bohmian mechanics. 
Below we show that this is no longer the case if the framework is extended properly, that is,  
we can actually regard Bohmian mechanics as an (extended) ontological model in which the QP is embedded.

\subsection{Ontological Models and Synlogicality}

For the extension, we first
introduce a certain type of contextuality and then formulate the ontological model based on
the QP.   Since our contextuality is slightly different from the contextuality discussed in \cite{Spekkens_Contextuality}, we employ
the term \lq synlogical\rq\ instead of \lq contextual\rq\ which was initially used in \cite{Kochen-Specker}. 

Let $\psi$ be a preparation of the system, whether or not it is classical or quantum, for which we measure an observable $A$. 
The preparation may be realized by preparing a state for the system, and in classical mechanics its complete specification is provided by a point $\gamma$ in phase space $\Gamma$ but more generally it is specified by a probability distribution $p\left(\gamma\left|\psi\right.\right)$ in  $\Gamma$.
Thus, in classical mechanics for which the physical reality is taken for granted, 
the probability distribution $p\left(a\left| \psi\right.\right)$ of obtaining the measurement result $a$ under the preparation $\psi$
can be written in the form,
\begin{equation}
p\left(a\left|\psi\right.\right)=\int_{\Gamma}p\left(a\left|\gamma,\psi\right.\right)p\left(\gamma\left|\psi\right.\right)\mathrm{d}\gamma.
\end{equation}

On the other hand, in quantum mechanics for which no physical reality analogous to the classical one is attached, the complete specification of preparation is provided by a
vector (pure state) in a Hilbert space $\mathcal{H}$.  Nevertheless one may consider
an ontological model which purports to describe experimentally observed phenomena supposing some physical reality, and such a model may be formulated as follows \cite{ontological_model_original, Spekkens_Contextuality}.
Let $\Lambda$ be a set of elements $\lambda$ which correspond to parameters (or hidden variables) representing the purported physical reality of the system.
We shall call the element $\lambda$ an \emph{ontic state} and the set $\Lambda$ an \emph{ontic state space}.
An ontological model is a model that is characterized by an ontic state $\lambda\in\Lambda$ and 
a (conditional) joint probability $p\left(\lambda,a\left|A,\psi\right.\right)$
of $\lambda$ with outcome $a\in\mathbb{K}_{A}$, 
where $\mathbb{K}_{A}$ is a set of all possible outcomes of the measurement of an observable $A$.
The argument $A$ in $p\left(\lambda,a\left|A,\psi\right.\right)$ might look redundant, but it proves important to keep track of what is measured in order to specify
the condition of measurement in the following argument.

Let $p\left(a\left|\psi\right.\right)$ be the probability distribution of the outcomes $a\in\mathbb{K}_{A}$ obtained
under the preparation $\psi$. 
From the conditional joint probability $p\left(\lambda,a\left|A,\psi\right.\right)$ provided in the model, the probability $p\left(a\left|\psi\right.\right)$ may be given  by
summing over all possible ontic states appearing in the preparation,
\begin{equation}
p\left(a\left|\psi\right.\right)=\int_{\Lambda}p\left(\lambda,a\left| A,\psi\right.\right)\mathrm{d}\lambda.
\label{eq:reproduction condition-1}
\end{equation}
We assume that in our ontological model the conditional joint probability always fulfills this \lq reproduction condition\rq\ (\ref{eq:reproduction condition-1}).
For our later purpose, we also extend the framework of the ontological model by allowing the conditional joint probability $p\left(\lambda,a\left|A,\psi\right.\right)$ to be
complex so that the QP may be admitted.  This extension does not affect the formal structure of the ontological model we are currently considering.

Given the conditional joint probability, the marginal probability of $\lambda$ is defined according to the standard procedure as
\begin{equation}
p\left(\lambda\left| A,\psi\right.\right):=\int_{\mathbb{K}_{A}}p\left(\lambda,a\left| A,\psi\right.\right)\mathrm{d}a,
\label{eq:marginal1}
\end{equation}
which has been referred to as {\it epistemic state} in \cite{ontological_model_original}.
Note that the marginal probability (\ref{eq:marginal1}) may not be uniquely determined from the preparation $\psi$ alone.
In fact, as the argument suggests, it could depend on the choice of the observable $A$. 
If, however, the marginal probability is independent of $A$ so that we can write
\begin{equation}
p\left(\lambda\left| A,\psi\right.\right)=p\left(\lambda\left| \psi\right.\right),
\label{eq:O-NS}
\end{equation}
for any $A$ and a pair of $\psi$ and $\lambda\in\Lambda$, then we call such a model \emph{observable-asynlogical}  (O-AS).
Otherwise, it is called \emph{observable-synlogical} (O-S).

Out of the joint
probability $p\left(\lambda,a\left| A,\psi\right.\right)$ and other marginal probabilities mentioned above, 
we define two types of conditional probabilities,
\begin{eqnarray}
p\left(a\left|\lambda,A,\psi\right.\right)
:=\frac{p\left(\lambda,a\left| A,\psi\right.\right)}{p\left(\lambda\left| A,\psi\right.\right)},
\qquad
p\left(\lambda\left|a,A,\psi\right.\right)
:=\frac{p\left(\lambda,a\left| A,\psi\right.\right)}{p\left(a\left|A,\psi\right.\right)},
\label{eq:conditional}
\end{eqnarray}
which are called {\it  indicator functions} in \cite{ontological_model_original}.  By construction, these functions fulfill Bayes' formula,
\begin{equation}
p\left(\lambda\left|a,A,\psi\right.\right)=\frac{p\left(a\left|\lambda,A,\psi\right.\right)p\left(\lambda\left| A,\psi\right.\right)}{p\left(a\left|A,\psi\right.\right)}.
\label{eq:Bayes}
\end{equation}

We also notice that the resultant conditional probabilities (\ref{eq:conditional}) depend on 
the preparation $\psi$ in general.  This alludes us to call an ontological model \emph{preparation-asynlogical} (P-AS)
if it is independent of $\psi$ allowing us to write
\begin{equation}
p\left(a\left|\lambda,A,\psi\right.\right)=p\left(a\left|\lambda,A\right.\right),
\label{eq:P-NC}
\end{equation}
for any $A$, $\lambda\in\Lambda$ and $a\in\mathbb{K}_{A}$. Otherwise, it is called \emph{preparation-synlogical} (P-S).

Using the materials we have introduced, the reproduction condition
(\ref{eq:reproduction condition-1}) can now be expressed as 
\begin{equation}
p\left(a\left|\psi\right.\right)=\int_{\Lambda}p\left(a\left|\lambda,A,\psi\right.\right)p\left(\lambda\left| A,\psi\right.\right)\mathrm{d}\lambda.\label{eq:reproduction2}
\end{equation}
If, in particular, the ontological model is both O-AS and P-AS, we have
\begin{equation}
p\left(a\left|\psi\right.\right)=\int_{\Lambda}p\left(a\left|\lambda, A\right.\right)p\left(\lambda\left|\psi\right.\right)\mathrm{d}\lambda.\label{eq:original_ontological_model}
\end{equation}
When an ontological model is of this type,  we shall simply call the model {\it asynlogical}, and otherwise we call it {\it synlogical}.
The conventional framework of the ontological (or hidden variable) model \cite{Bell1964, ontological_model_original} is confined to the asynlogical case, but below we need to deal with the synlogical case in order to accommodate Bohminan mechanics in the framework.

\subsection{Bohminan Mechanics as a Synlogical Ontological Model}

Now we show that Bohmian mechanics is a quasiprobabilistic
P-S ontological model.  To this end, we first recall that the ontic state space of Bohmian mechanics
is just the position eigenspace,
\begin{equation}
\Lambda=\left\{ \left|\mathbf{x}\right\rangle \left|\, \mathbf{x}\in\mathbb{R}^{N}\right.\right\},
\label{eq: bohmontsp}
\end{equation}
where $N = 3n$ if $n$ particles are present in the three dimensional space.
Next, we observe that, in view of (\ref{eq:extc2}), the QP underlying condition B$3^\prime$ is 
the indicator function $p^{\alpha}\left(a\left|\psi,\mathbf{x}\right.\right)$ given in (\ref{eq:conditionalquasipro. dBB}), or 
the joint QP in (\ref{eq:dBBjointQuasi}) from which the Born rule follows directly as noted earlier.
Combining these, we find that Bohmian mechanics can actually be regarded as an 
ontological model defined by the ontic state space $\Lambda$ in (\ref{eq: bohmontsp}) and the joint QP $p^{\alpha}\left(a\left|\psi,\mathbf{x}\right.\right)$
in (\ref{eq:dBBjointQuasi}).
Since $p^{\alpha}\left(a\left|\psi,\mathbf{x}\right.\right)$
depends on $\psi$, Bohmian mechanics, regarded this way, is a P-S ontological model.

Having found the intrinsic QP associated with Bohmian mechanics, we now proceed backwardly and deduce 
the basic ingredients of Bohmian mechanics from the QP.  First, from 
the QP the epistemic state (\ref{eq:marginal1}) is given by
\begin{equation}
\int p^{\alpha}\left(\mathbf{x},a\left|\psi\right.\right)\mathrm{d}a=p(\mathbf{x}|\psi)=\left|\left\langle \mathbf{x}|\psi\right\rangle \right|^{2},
\end{equation}
which is just condition B2.  Second, the indicator functions (\ref{eq:conditional}) imply 
\begin{equation}
p^{\alpha}\left(a\left|\mathbf{x}, \psi\right.\right)=\frac{p^{\alpha}\left(\mathbf{x},a\left|\psi\right.\right)}{p(\mathbf{x}|\psi)},
\qquad
p^{\alpha}\left(\mathbf{x}\left| a, \psi\right.\right)=\frac{p^{\alpha}\left(\mathbf{x}, a\left|\psi\right.\right)}{p\left(a\left|\psi\right.\right)},
\end{equation}
and, accordingly, the physical value corresponding to the observable $A$ takes the form,
\begin{eqnarray}
\left\langle A\right\rangle_\psi^{\alpha}\left(\mathbf{x}\right) 
&=& \int a\, p^{\alpha}\left(a\left|\mathbf{x}, \psi\right.\right)\mathrm{d}a \nonumber \\
&=& \alpha\frac{\left\langle \mathbf{x}\right|A\left|\psi\right\rangle }{\left\langle \mathbf{x}|\psi\right\rangle }
 +\left(1-\alpha\right)\frac{\left\langle \psi\right|A\left|\mathbf{x}\right\rangle }{\left\langle \psi|\mathbf{x}\right\rangle},
\label{eq:defpval}
\end{eqnarray}
which is just condition B$3'$.

Proceeding a step further, let us consider the question of quantum nonlocality in Bohmian mechanics.  For this, we need to examine the correlation of measurement outcomes of two observables, $A$ and $B$, which may be evaluated by a straightforward extension of (\ref{eq:defpval}):
\begin{eqnarray}
\left\langle A\, B \right\rangle^\alpha_{\psi} \left(\mathbf{x}\right) 
&=& \int ab \, p^{\alpha}\left(a, b\left|\mathbf{x}, \psi\right.\right)\mathrm{d}a\, \mathrm{d}b,
\label{eq:correl}
\end{eqnarray}
using the conditional probability,
\begin{equation}
p^{\alpha}\left(a, b\left|\mathbf{x}, \psi\right.\right)=\frac{p^{\alpha}\left(\mathbf{x},a, b\left|\psi\right.\right)}{p(\mathbf{x}|\psi)},
\end{equation}
defined from the joint probability,
\begin{equation}
p^{\alpha}\left(\mathbf{x},a, b\left|\psi\right.\right)=\left\langle \psi\right|E^{X}\left(\mathbf{x}\right)\circ_{\alpha}\left(E^{A}\left(a\right)E^{B}\left(b\right)\right)\left|\psi\right\rangle.
\end{equation}

For definiteness, let us consider two particles labeled by 1 and 2 with position $\mathbf{x}_1$ and $\mathbf{x}_2$ for which we measure $A$ for particle 1 and $B$ for particle 2, respectively.  First, if the observables $A$, $B$ commute with the positions $\mathbf{x}_1, \mathbf{x}_2$, as in the case where, {\it e.g.}, we measure the spins of respective particles, then 
the $\alpha$-dependence disappears as we mentioned earlier.  Moreover, if the state $\left|\psi\right\rangle$ is a direct product of the states of the two particles, $\left|\psi\right\rangle = \left|\psi_1\right\rangle\left|\psi_2\right\rangle$, then, noting $E^{X}\left(\mathbf{x}\right) = E^{X}\left(\mathbf{x}_1\right) E^{X}\left(\mathbf{x}_2\right)$, we have (suppressing $\alpha$ which does not appear in this case),
\begin{eqnarray}
p\left(\mathbf{x},a, b\left|\psi\right.\right)
&=& \left\langle \psi_1\right|E^{X}\left(\mathbf{x}_1\right)E^{A}\left(a\right)\left|\psi_1\right\rangle
\left\langle \psi_2\right|E^{X}\left(\mathbf{x}_2\right)E^{B}\left(b\right)\left|\psi_2\right\rangle \nonumber \\
&=& p\left(\mathbf{x}_1,a\left|\psi_1\right.\right)p\left(\mathbf{x}_2,b\left|\psi_2\right.\right).
\label{eq:localdecomp}
\end{eqnarray}
Using the relation,
\begin{equation}
p(\mathbf{x}|\psi)=\left|\left\langle \mathbf{x}|\psi\right\rangle \right|^{2} 
= \left|\left\langle \mathbf{x}_1|\psi_1\right\rangle \right|^{2} \left|\left\langle \mathbf{x}_2|\psi_2\right\rangle \right|^{2}
= p(\mathbf{x}_1\left|\psi_1\right.) \, p(\mathbf{x}_2\left|\psi_2\right.),
\end{equation}
we see that in this particular case,
\begin{eqnarray}
\left\langle A\, B \right\rangle_{\psi} \left(\mathbf{x}\right) 
&=& \int a\, p(a\left|\mathbf{x}_1, \psi_1\right.)\, b\, p(a\left|\mathbf{x}_2, \psi_2\right.)\, \mathrm{d}a \mathrm{d}b \nonumber \\
&=& \left\langle A\right\rangle_{\psi_1}
\left(\mathbf{x}_1\right) \left\langle B\right\rangle_{\psi_2}\left(\mathbf{x}_2\right),
\label{eq:correltwo}
\end{eqnarray}
where
\begin{equation}
\left\langle A\right\rangle_{\psi_1}\left(\mathbf{x}_1\right) =  \frac{\left\langle \mathbf{x}_1\right|A\left|\psi_1\right\rangle }{\left\langle \mathbf{x}_1|\psi_1\right\rangle},
\qquad
\left\langle B\right\rangle_{\psi_2}\left(\mathbf{x}_2\right) =  \frac{\left\langle \mathbf{x}_2\right|A\left|\psi_2\right\rangle }{\left\langle \mathbf{x}_2|\psi_2\right\rangle},
\end{equation}
which coincide with the local expectation value (\ref{eq:LEV}) for the respective particles.
It is evident from the resultant product structure in (\ref{eq:correltwo}) that the correlation is local, which is to be compared with the familiar expression of the value in the hidden variable theory \cite{Bell_HVT}.  
On the other had, the above argument implies that the correlation may become nonlocal if the state is not a direct product
$\left|\psi\right\rangle \ne \left|\psi_1\right\rangle\left|\psi_2\right\rangle$.

If the observables $A$, $B$ do not commute with the positions $\mathbf{x}_{1}$, $\mathbf{x}_{2}$, as in the case where, {\it e.g.}, we measure the angular momenta of the respective particles, then we see that 
the $\alpha$-dependence remains intact in general even for the product state.  Indeed, in this setting we have
\begin{eqnarray*}
p^{\alpha}\left(a,b\left|\mathbf{x},\psi\right.\right) 
& = & \alpha\frac{\left\langle \mathbf{x}\right|E^{A}\left(a\right)E^{B}\left(b\right)\left|\psi\right\rangle }{\left\langle \mathbf{x}|\psi\right\rangle }+\left(1-\alpha\right)\frac{\left\langle \psi\right|E^{A}\left(a\right)E^{B}\left(b\right)\left|\mathbf{x}\right\rangle }{\left\langle \psi|\mathbf{x}\right\rangle } \nonumber \\
& = & 
\alpha\frac{\left\langle \mathbf{x}_{1}\right|E^{A}\left(a\right)\left|\psi_{1}\right\rangle }{\left\langle \mathbf{x}_{1}|\psi_{2}\right\rangle }\frac{\left\langle \mathbf{x}_{2}\right|E^{B}\left(b\right)\left|\psi_{2}\right\rangle }{\left\langle \mathbf{x}_{2}|\psi_{2}\right\rangle }+\left(1-\alpha\right)\frac{\left\langle \psi_{2}\right|E^{A}\left(a\right)\left|\mathbf{x}_{1}\right\rangle }{\left\langle \psi_{2}|\mathbf{x}_{1}\right\rangle }\frac{\left\langle \psi\right|E^{A}\left(a\right)\left|\mathbf{x}_{2}\right\rangle }{\left\langle \psi_{2}|\mathbf{x}_{2}\right\rangle }.
\end{eqnarray*}
This implies that the correlation is nonlocal even for product states, except for the two trivial cases $\alpha = 1$ and $0$ for which we find
\begin{eqnarray}
p^{\alpha=1}\left(a,b\left|\mathbf{x},\psi\right.\right)
& = & p^{\alpha=1}\left(a\left|\mathbf{x}_{1},\psi_{1}\right.\right)p^{\alpha=1}\left(b\left|\mathbf{x}_{2},\psi_{2}\right.\right), \nonumber \\
p^{\alpha=0}\left(a,b\left|\mathbf{x},\psi\right.\right)
& = & p^{\alpha=0}\left(a\left|\mathbf{x}_{1},\psi_{1}\right.\right)p^{\alpha=0}\left(b\left|\mathbf{x}_{2},\psi_{2}\right.\right),
\end{eqnarray}
or equivalently,
\begin{eqnarray}
\left\langle A\, B \right\rangle_{\psi}^{\alpha=1} \left(\mathbf{x}\right) 
&=& \left\langle A\right\rangle_{\psi_1} \left(\mathbf{x}_1\right) \left\langle B\right\rangle_{\psi_2}\left(\mathbf{x}_2\right), \nonumber \\
\left\langle A\, B \right\rangle_{\psi}^{\alpha=0} \left(\mathbf{x}\right) 
&=& \left\langle A\right\rangle_{\psi_1}^* \left(\mathbf{x}_1\right) \left\langle B\right\rangle_{\psi_2}^*\left(\mathbf{x}_2\right).
\end{eqnarray}

An alternative, reasonable definition of the joint QP for the composite
system may be given by
\begin{eqnarray}
p^{\alpha_1\alpha_2}\left(\mathbf{x},a, b\left|\psi\right.\right)
=\left\langle \psi\right|E^{A}\left(a\right)\circ_{\alpha_1}E^{X_{1}}\left(\mathbf{x}_{1}\right)E^{B}\left(b\right)\circ_{\alpha_2}E^{X_{2}}\left(\mathbf{x}_{2}\right)\left|\psi\right\rangle,
\end{eqnarray}
where $\alpha_1$ and $\alpha_2$ are the complex parameters specifying the ambiguity of joint QP assigned to the respective two particles.
It is clear that this confines the ambiguity within each of the particles and, accordingly as in (\ref{eq:localdecomp}), it ensures locality in the correlation for all $\alpha_1$ and $\alpha_2$ 
for any product states $\left|\psi\right\rangle =\left|\psi_{1}\right\rangle \left|\psi_{2}\right\rangle $,
\begin{eqnarray}
p^{{\alpha_1\alpha_2}}\left(\mathbf{x},a, b\left|\psi\right.\right)
=p^{\alpha_1}\left(\mathbf{x}_{1}, a\left|\psi_{1}\right.\right)p^{\alpha_2}\left(\mathbf{x}_{2}, b\left|\psi_2\right.\right).
\end{eqnarray}

Before closing this section, we remark that, even when the preparation $\psi$ is described by
a mixed state $\rho$ in the foregoing arguments, the corresponding result is obtained simply by replacing $\left\langle \psi\right|\ldots\left|\psi\right\rangle$
with $\mathrm{Tr}\left[\dots\rho\right]$. 
Incidentally, we also remark that since there are vectors $\psi,\phi\in\mathcal{H}$ such that 
\begin{equation}
p(\mathbf{x}|\psi)\, p\left(\mathbf{x}\left|\phi\right.\right)=\left|\left\langle \mathbf{x}|\psi\right\rangle \right|^{2}\left|\left\langle \mathbf{x}|\phi\right\rangle \right|^{2}\neq 0,
\end{equation}
Bohmian mechanics is \lq $\psi$-epistemic\rq\ according to the classification advocated in \cite{ontological_model_original,Spekkens_Contextuality}.

\section{Conclusion and Discussion}

In this paper we have argued, from both the structural and the conceptual viewpoints, that the conditional QP defined from the weak value provides 
a basic ingredient of quantum theory.  
In the structural viewpoint, we have found that the QP forms the fundamental element in the description of a quantum transition process, and that 
it is characterized by a complex $\alpha$-parameter representing the ambiguity of the process.  Curiously, or reassuredly,
this ambiguity disappears at the final stage of dealing with physically observable quantities.  

In the conceptual viewpoint, we have seen  
that the QP can be embedded in Bohmian mechanics such that one of the postulates of Bohmian mechanics is replaced by an alternative one from which the Born rule is derived directly.   This observation allows us to recognize 
Bohmian mechanics as an ontological model of a certain synlogical (contextual) type, clarifying its so far obscure status in the category of hidden variable models.

It should be stressed that the QP introduced in this paper is defined for two non-commuting observables $A$ and $B$ for which
no joint probability is admitted in quantum mechanics on account of the incompatibility of simultaneous measurements of the two observables.
Compared to the QP discussed earlier ({\it e.g.}, \cite{Wigner_Function,Kirkwood,Ozawa_QuasiProb}), our QP possesses 
the $\alpha$-dependence which   
specifies the degree of mixture of transition processes under time reversal.  
In the practical side, our QP serves as a useful tool for treating 
statistical quantities such as the expectation value and the correlation of two physical observables in a manner analogous to the conventional probability.
This will be particularly convenient for dealing with situations where such expectation values and correlations take non-standard values, which will occur when the weak value is considered in specific arrangements to achieve amplification \cite{Aharonov1} for instance.

\begin{acknowledgements}
We thank R. Koganezawa for his valuable suggestion in the early stage of the work.
This work was supported in part by JSPS KAKENHI No.~25400423,  No.~26011506, and by the Center for the Promotion of Integrated Sciences (CPIS) of SOKENDAI.
\end{acknowledgements}

\appendix*

\section{Some useful Formulas for the $\circ_{\alpha}$-Product}

The $\circ_{\alpha}$-product (\ref{eq:alphaproduct}) for two operators $X$ and $Y$
on $\mathcal{H}$ is defined by  
\begin{equation}
X\circ_{\alpha}Y:=\alpha XY+\left(1-\alpha\right)YX,
\end{equation}
with $\alpha\in\mathbb{C}$. 
For $\alpha=1$ and $\alpha=0$, the $\circ_{\alpha}$-product becomes
the usual operator product, 
\begin{equation}
X\circ_{\alpha=1}Y=XY, \qquad 
X\circ_{\alpha=0}Y=YX,
\end{equation}
whereas for $\alpha=1/2$ it reduces to the Jordan product \cite{Jordan_product},
\begin{equation}
X\circ_{\alpha=\frac{1}{2}}Y=\frac{1}{2}\left(XY+YX\right).
\end{equation}

If we put $\alpha = s + it$ with real $s, t$, we have 
\begin{equation}
X\circ_{\alpha=s+it}Y =  s XY+\left(1-s\right)YX+it \left[X,Y\right] =  X\circ_s Y+it\left[X,Y\right],
\end{equation}
where $[X,Y]=XY-YX$ is the commutator.
In particular, for $\alpha=\frac{1}{2}\left(1-i\right)$ we find
\begin{equation}
X\circ_{\alpha=\frac{1}{2}\left(1-i\right)}Y=\frac{1}{2}\{X,Y\}+\frac{1}{2i}[X,Y],
\end{equation}
where $\{X,Y\}=XY+YX$ is the anti-commutator.
In addition, the $\circ_{\alpha}$-product has the following properties:
\begin{eqnarray}
I\circ_{\alpha}X & = &X\circ_{\alpha}I=X, 
\label{eq:xident} \\
X\circ_{\alpha}X & = &X^{2}, \\
X\circ_{\alpha}Y-Y\circ_{\alpha}X & = &\left(2\alpha-1\right)[X,Y],
\label{eq:commutatorrrrr} \\
X\circ_{\alpha}Y+Y\circ_{\alpha}X & = & XY+YX, \\
{} [X , Y]  = 0\, & \Rightarrow& \,  X \circ_{\alpha} Y=XY=YX,
\end{eqnarray}
with $I$ being the identity operator on $\mathcal{H}$.



\end{document}